# A novel conservative chaos driven dynamic DNA coding for image encryption

**Vinod Patidar**[1,*] and **Gurpreet Kaur**[2]

[1]School of Engineering
Sir Padampat Singhania University
Bhatewar, Udaipur-313601, Rajasthan, India
email: vinod.patidar@spsu.ac.in
ORCID ID: 0000-0002-1270-3454

[2]Amity Institute of Information Technology
Amity University
Noida-201313, UP, India
email: gkaur10@amity.edu
ORCID ID: 0000-0002-2611-5143

**Abstract:**
Recently, many image encryption algorithms based on hybrid DNA and chaos have been developed. Most of these algorithms utilize the chaotic systems with dissipative dynamics and exhibiting periodic windows and patterns in bifurcation diagrams along with co-existing attractors in the neighbourhoods of parameter space. Therefore, such algorithms generate several weak keys, thereby making them prone to various chaos- specific attacks.  In this paper, we propose a novel conservative chaotic standard map-driven dynamic DNA coding (encoding, addition, subtraction and decoding) for the image encryption. The proposed image encryption algorithm is a dynamic DNA coding algorithm i.e., for the encryption of each pixel different rules for encoding, addition/subtraction, decoding etc. are randomly selected based on the pseudorandom sequences generated with the help of the conservative chaotic standard map.  We propose a novel way to generate pseudo-random sequences through the conservative chaotic standard map and also test them rigorously through the most stringent test suite of pseudo-randomness, the NIST test suite, before using them in the proposed image encryption algorithm. Our image encryption algorithm incorporates a unique feed-forward and feedback mechanisms to generate and modify the dynamic one-time pixels that are further used for the encryption of each pixel of the plain image, therefore, bringing in the desired sensitivity on plaintext as well as ciphertext. All the controlling pseudorandom sequences used in the algorithm are generated for a different value of the parameter (part of the secret key) with inter-dependency through the iterates of the chaotic map (in the generation process) and therefore possess extreme key sensitivity too. The performance and security analysis has been executed extensively through histogram analysis, correlation analysis, information entropy analysis, DNA sequence-based analysis, perceptual quality analysis, key sensitivity analysis, plaintext sensitivity analysis, etc., The results are promising and prove the robustness of the algorithm against various common cryptanalytic attacks.

___________________

*Corresponding Author

# 1. Introduction

Due to advancements in network and communication technologies, the exchange of digital multimedia content has become one of the frequent tasks. It has consequently posed a requirement to protect such digital multimedia information from eavesdropping. Amongst various digital multimedia contents, images (and hence videos too) require special attention due to some of the inherent properties of digital images like the bulk of information, high spatial correlation and redundancies. Consequently, over the years, the encryption of images has been one of the active areas of research in image processing and allied fields and therefore a variety of technologies like optical image encryption, chaos-based image encryption, DNA based encryption, and a suitable combination of these technologies have emerged as alternative means to encrypt the images.

Claudio Shanon [1] in his masterpiece "Communication theory of secrecy systems," suggested that good mixing transformations governed by simple repeating and non-commuting nonlinear operations involving the secret key in a complex way are the key ingredients for developing an ideal encryption system. Such transformations may be comfortably realized through the confusion (substitution) and diffusion (permutation) mechanisms. Here the confusion means a complex and involved relationship between the cipher image pixels and key whereas diffusion refers to spreading the plain image pixels information over the entire cipher image.

The optical image encryption systems utilize optical setups (lenses, spatial lens modulators, etc.) [2-4], double random phase encoding (DRPE) with optically or electronically generated random phase masks [4, 5-8], and mathematical modelling with integral transforms [4, 9-10]. Such systems have the advantage of sending complex data in parallel and are also capable of carrying out usually time-consuming operations in a faster way, therefore are found suitable in image encryption. Besides having above mentioned clear advantages, the optical processes governed by the integral transforms possess the linearity and symmetry properties which make the optical encryption system vulnerable to various cryptanalytic attacks [4, 11-13]. On the other hand, DNA computing, since its advent in 1994 [14], has attracted the attention of researchers due to some of its peculiar features like huge information-carrying capacity, parallelism, ultra-low energy consumption etc. DNA computing mainly requires the biochemical reaction environment, expensive laboratory equipment, and restricted laboratory conditions like precise control of concentration, temperature and pH of biochemical reactants etc., which make it difficult to realize in a wet lab. Rather than implementing it at a molecular level, researchers have preferred DNA coding to carry the information in digital form and manipulate it using the corresponding feasible DNA operations. It has induced a new way of concealing the information through DNA microdots [15] and subsequently following this development, Gehani [16], Xiao [17] and Kang [18] too presented the new perspectives of information hiding using the DNA concepts. Optical transforms and DNA encoding/operations do not offer nonlinearity therefore solely are not suitable to develop secure encryption systems as per Shanon's criterion. Contemporary to the above-mentioned developments of optical and DNA-based image encryption, the dynamical chaos has also been extensively used to develop secure image encryption systems owing to the fact that chaotic systems are essentially nonlinear systems (having sensitivity on initial conditions/parameters, ergodicity, mixing property etc.) and have been found most suitable to introduce the substitution and permutation of image pixels as recommended in Shannon's confusion-diffusion framework [19-21]. Chaos-based image encryption systems, have also been preferred due to their fast-processing time which is one of the essential requirements in real-time transmission. However, there are some limitations associated with chaos-based encryption systems like smaller key space, floating-point representation, periodic windows and patterns in the bifurcation diagram, coexisting attractors in the neighbourhoods of parameter space etc. [22, 23].

Since each of the technologies, mentioned above has its inherent advantages as well as disadvantages, therefore, researchers find it worthwhile to hybridize various techniques in order to either incorporate each of their pros or eliminate any of their cons. In such hybrid methods, the chaotic dynamical systems have been mainly utilized to introduce the nonlinear effects in the substitution and permutation of pixels in a variety of ways. On the other hand, the optical transforms OR DNA operations have been utilized to do encoding/decoding of image pixels that too sometimes under the control of chaotic systems [4]. As the present manuscript deals with hybrid chaos and DNA based image encryption, therefore we are elaborating more on this category in our further discussion.

In hybrid DNA and chaos-based image encryption systems, the images are firstly encoded into the DNA sequences followed by scrambling of these sequences using chaotic systems (one dimensional, combination of multiple chaotic maps, hyperchaotic maps, combination of hyperchaotic maps). The DNA bases of scrambled sequences are then changed by the application of DNA operations (addition, subtraction, XOR, XNOR, DNA complements, or combinations of some of these operations etc.) under the control of chaotic systems and then the resultant sequences are decoded into the digital format to produce the encrypted image. Broadly the DNA and chaos-based image encryption can be classified into the following categories [24]: (i) fixed DNA coding, (ii) dynamic DNA coding, (iii) DNA base complement operations, (iv) DNA sequence algebraic operations, and (v) combinations of multiple DNA operations. In the *fixed DNA coding schemes,* a particular rule is used for encoding followed by some DNA operations and decoding using the same rule [25-29], however in the *dynamic DNA coding schemes*, different rules are used for the encoding and decoding (either row-wise, column-wise, block-wise, pixel-wise or sometimes at the base-wise too) under the control of chaotic system [30-34]. In *DNA base compliment operations schemes*, one of the three types of complement methods (single base direct complement method, static regular base complement method, dynamic regular base complement method) are used [32, 34-37]. However, in *DNA sequence algebraic operations-based schemes*, addition, subtraction, XOR, XNOR operations on DNA sequences are used [38-41] under the control of chaotic systems to change the pixel values. In the last category, combined DNA coding and multiple/different DNA operations are used to scramble and change the pixel values [36, 39] and therefore are the most complex scenario. We are reviewing some of the very important and recent research works which have paved the way and led to the advancements in this field of hybrid DNA and chaos-based image encryption algorithms.

Zhang et al. [35] proposed an image encryption scheme based on DNA addition operation in which a DNA encoded image is divided into blocks and DNA addition is used to add these blocks followed by a complement operation with the help of the chaotic logistic map. A 4D hyperchaotic map is used to generate pseudo-random number sequences and a circular permutation along with classical diffusion is used for the encryption [42]. Liu et al. [28] also used DNA addition and complement to develop the image encryption of RGB images. Wei et al. [27] used Chen's hyperchaotic system to scramble the pixels of RGB layers and then divide them into some equal blocks followed by the addition of these blocks under the influence of Chen's hyperchaotic system. A confusion-diffusion based image encryption based on a piecewise linear chaotic map and Chebyshev map and DNA complementary rules has been proposed by Liu et al.[43]. Enayatifar et al. [44] proposed a hybrid genetic algorithm which is used for determining the best DNA mask out of several such masks generated through the chaotic system and then further used for image encryption. Wu et al. [45] proposed a new robust color image encryption scheme based on dynamic DNA sequence operations and multiple improved/compound 1D chaotic systems which utilizes a division shuffling process and the key streams are dependent on the secret key and plaintext. Kalpana and Murali [30] introduced the concept of using more than one DNA rule and more than one operation (subtraction/addition) in the algorithm which is randomly chosen for each pixel with the help of multiple chaotic systems such as Chen's hyperchaotic map, sine map, cubic map, logistic map and Arnold's chaotic maps have been used. A chaotic logistic map and spatiotemporal system (coupled map lattices) are used in combination with the DNA rules to achieve the image encryption system through a permutation-substitution architecture [31]. Chai et al. [46] proposed an image encryption algorithm using the 2D logistic map to execute the row and column circular permutations where SHA256 is used to generate the initial conditions for chaotic maps. A new DNA coding of images along with two rounds of DNA-based confusion and diffusion, where a piecewise linear chaotic map is used to generate the key stream, is proposed by Zhang [47]. A combined Block-based permutation, pixel-based substitution, DNA encoding, bit-level substitution (i.e., DNA complementing), DNA decoding, and bit-level diffusion are used for image encryption where the logistic-Chebyshev map, sine-Chebyshev map produces the key-streams at various stages given above [48]. Chai et al. [49] proposed a novel diffusion mechanism based on random numbers related to plaintext (DMRNRP) is used along with DNA operations under the control of a four wings hyperchaotic system. A one-time pad colour image encryption based on a 3D skew tent map utilizing the secret keys and Hamming distance is proposed in which DNA XOR, addition and subtraction are used [50]. Wang et al. [34] proposed a one-time pad image encryption algorithm based on the coupled map lattices (CML system) and DNA diffusion sequences. The initial values and control parameters of the CML system and logistics map serve as keys for a one-time pad and are calculated by utilizing the SHA256 hash algorithm. DNA encoding along with other operations is used at the base level under the influence of chaos. A new scheme was proposed by combining the optimal coding mechanism with the optimal DNA coding operation [24]. Another one-time pad DNA-chaos image encryption algorithm, based on multiple keys and utilizing the chaotic logistic and sine maps, is proposed by Zhou [51] in which plaintext sensitivity is integrated by having dependence of four of the keys on the original image. A robust medical image encryption based on a combined DNA-chaos approach for secure telemedicine utilizing the logistic map, piecewise linear chaotic map (PWLCM), DNA encoding and various DNA algebraic operations like XOR, addition, subtraction etc. for the diffusion [52]. Wang et al. [53] proposed an image encryption strategy based on random number embedding in the plaintext and DNA-level self-adaptive permutation and diffusion based on a 4D memristive hyperchaotic system. A new four-dimensional hyperchaotic system is proposed by Hui et al. [54] and used further to encrypt the original image through pixel scrambling and pixel diffusion based on DNA encoding. A chaotic logistic map-based image encryption algorithm utilizing the arithmetic sequence model scrambling method and DNA operations is proposed by Yan et al. [55]. An efficient DNA-inspired image encryption algorithm based on the fusion of hyper-chaotic diffusion and wavelet-based confusion is proposed by El-Khamy and Mohamed [56]. In Table 1, we have summarized the recent algorithm that are most relevant to present work along with their characteristic components and performance metrics.

**Table 1:** Summary of recent hybrid DNA and chaos-based image encryption algorithms

| Reference | Chaotic system | Use of Hash | Key space | Correlation | | | NPCR | UACI |
|---|---|---|---|---|---|---|---|---|
| | | | | H | V | D | | |
| [35] | Logistic | - | $10^{72}$ | 0.0036 | 0.0023 | 0.0039 | 99.61 | 0.38 |
| [28] | Logistic | - | $10^{56}$ | 0.0059 | −0.0042 | 0.0180 | - | - |
| [28] | Chen's | - | $10^{70}$ | 0.0054 | 0.0016 | 0.0017 | 99.58 | 33.48 |
| [43] | PWLCM, Chebyshev | MD5 | $10^{126}$ | 0.0004 | 0.0021 | −0.0038 | 99.53 | 32.57 |
| [44] | Logistic | - | | 0.0017 | 0.0007 | 0.0001 | 99.71 | 33.62 |
| [45] | Logistic-tent and Logistic-sine | - | $10^{90}$ | −0.0084 | 0.0004 | −0.0015 | 99.60 | 33.48 |
| [30] | Lorenz, Chen's hyperchaos, Logistic, Sine, Cubic, Arnold's | - | $10^{230}$ | 0.0144 | 0.0083 | −0.0467 | 99.65 | 33.11 |
| [31] | Logistic (Coupled map lattice) | - | $10^{89}$ | 0.0214 | 0.0465 | −0.0090 | 99.60 | 33.44 |
| [46] | Logistic | SHA256 | $>2^{100}$ | −0.0045 | $-1.62 \times 10^{-4}$ | 0.0053 | 99.59 | 33.41 |
| [47] | PWLCM | - | $2^{16d}$ (d=28/64) | −0.0052 | 0.0221 | −0.0103 | 99.61 | 33.47 |
| [48] | Logistic-Chebyshev | SHA256 | $2^{716}$ | 0.0013 | −0.0049 | 0.0057 | 99.65 | 33.41 |
| [49] | 4-wing hyperchaos | SHA384 | $10^{185}$ | −0.0029 | 0.0013 | 0.0004 | 99.61 | 33.50 |
| [50] | Skew tent map, Coupled map lattice | - | $>10^{150}$ | −0.0085 | −0.0008 | 0.0033 | 99.60 | 33.43 |
| [51] | Sine, Chebyshev | - | $10^{112} \times 2^{25}$ | 0.0105 | −0.0025 | 0.0003 | 99.60 | 33.36 |
| [53] | Memristor hyperchaos | - | $2^{186}$ | 0.0084 | −0.0039 | −0.0013 | 99.62 | 30.64 |
| [54] | Lorentz based hyperchaos | SHA512 | $10^{60}$ | −0.0073 | −0.0042 | 0.0049 | 99.64 | 33.38 |
| [55] | Logistic | SHA512 | $2^{592} \times 10^{84}$ | 0.0023 | −0.0018 | 0.0013 | 99.61 | 33.54 |
| [56] | Chen's hyperchaos | - | - | 0.0009 | 0.0021 | 0.0003 | 99.61 | 33.43 |
| Proposed | Conservative Chaotic Standard map | | Infinite | 0.0003 | −0.0083 | 0.0007 | 99.61 | 33.51 |

It is evident from the review of hybrid DNA chaos-based image encryption algorithms that almost all of them are based on the chaotic logistic maps, sine map, cubic map, Arnold map, piecewise linear chaotic maps, their compound higher dimensional versions, Lorenz system, Rossler systems, 4D hyperchaotic systems like Chen's system and some of the newly developed hyperchaotic systems. In all such hybrid algorithms, chaotic systems are either used to control the substitution and permutation (DNA coding, encoding and algebraic operations) through the pseudo-random sequences generated by chaotic systems and/or to generate the one-time pads for further DNA based coding and operations to be used in the encryption. The chaotic systems used in all these algorithms are dissipative chaotic systems and exhibit several periodic windows and patterns in bifurcation diagrams and co-existing attractors in the neighbourhoods of parameter space and therefore possess several weak keys. Moreover, the processes of generating pseudorandom sequences, which are mainly controlling these algorithms and responsible for the nonlinearity in the algorithms, have not been rigorously tested for their pseudo-randomness. Consequently, such algorithms may be prone to chaos-based analysis/attacks.

To counter such possibilities, we propose a novel combination of conservative chaotic standard map-driven dynamic DNA encoding/decoding and operations (addition/subtractions) for image encryption. The conservative chaos map used in the proposed image encryption algorithm is a 2D map which exhibits robust chaos for all parameter values above a threshold (critical) value, and there exists no co-existing attractor too therefore, the chaotic orbit visits the entire phase space ergodically. Such ergodic orbits are highly recommended and proven best for the generation of pseudo-random sequence generations [57, 58]. We also propose a novel way to generate pseudo-random sequences (to be used in the proposed image encryption algorithm) through the conservative chaotic standard map and also test them rigorously through the most stringent test suite of pseudo randomness, the NIST test suite [59], by following all the recommendations of the test suite before using them in the proposed algorithm. Our image encryption algorithm incorporates a unique feed-forward and feedback mechanisms to generate and modify the dynamic one-time pixels that are further used for the encryption of each pixel of the plain image, thereby bringing in the desired extreme sensitivity on plaintext as well ciphertext. All the pseudo-random sequences used in the proposed image encryption algorithm are generated for an independent value of the parameter (part of the secret key) of the chaotic map and also have inter-dependency through the iterates of the chaotic map (in the generation process) therefore, the entire proposed algorithm possesses the extreme key sensitivity too. The complete details of the proposed image encryption algorithm have been described, in detail, in the next section.

## 2. The proposed image encryption algorithm

In this section, we describe the DNA coding/encoding and corresponding addition and subtraction operations being used in the proposed image encryption algorithm, the novel way of generating pseudorandom sequences based on a conservative chaotic standard map and their testing with the NIST pseudo-randomness test suite, and the finer algorithmic step-by-step details of the proposed image encryption algorithm and the entire flow of the encryption process.

**2.1 The DNA encoding/decoding and corresponding operations:**

The DNA sequences are comprised of four nucleic acid bases: Adenine (A), Thymine (T), Cytosine (C) and Guanine (G), here A and T are complements of each other and G and C are complements of each other. In DNA computing, these four nucleic bases are represented by 00, 01, 10 and 11. A total of 24 different combinations are possible for such representations out of which only 8 are allowed according to the complementarity rules of binary numbers (00 and 11 are complements, 01 and 10 are complements) and consistent with the DNA complement rule too. In Table 2, all 8 allowed representations or DNA encoding/decoding rules for binary numbers 00, 01, 10, and 11 have been depicted. In 8-bit image representation, each pixel value lies between 0 to 255 i.e., its binary representation is an 8-bit code therefore a pixel is represented by a combination of four nucleic bases. Therefore, a pixel can be encoded in eight different ways by following the DNA encoding rules given in Table 2 [34].

**Table 2:** DNA encoding/decoding rules

| Rule | 1 | 2 | 3 | 4 | 5 | 6 | 7 | 8 |
|---|---|---|---|---|---|---|---|---|
| 00 | A | A | T | T | C | C | G | G |
| 01 | G | C | G | C | A | T | A | T |
| 10 | C | G | C | G | T | A | T | A |
| 11 | T | T | A | A | G | G | C | C |

For each DNA encoding rule, the operations like addition, subtraction, XOR, XNOR, etc. can be defined by following the corresponding binary operation rules. Therefore, there are different rules for these operations corresponding to each DNA encoding rule. In the present algorithm, we are using addition and subtraction operations along with the DNA coding/encoding. For brevity, only addition and subtraction tables (Tables 3 and 4) corresponding to one of the DNA encoding/decoding rules (for rule no. 4) are provided here. Similarly, the addition/ subtraction tables for the remaining rules can also be developed.

**Table 3:** DNA addition rules (corresponding to DNA encoding/decoding rule no. 4)

| + | A | T | C | G |
|---|---|---|---|---|
| A | G | A | T | C |
| T | A | T | C | G |
| C | T | C | G | A |
| G | C | G | A | T |

**Table 4:** DNA subtraction rules (corresponding to DNA encoding/decoding rule no. 4)

| - | A | T | C | G |
|---|---|---|---|---|
| A | T | A | G | C |
| T | C | T | A | G |
| C | G | C | T | A |
| G | A | G | C | T |

In the proposed image encryption, we use any one of all eight DNA encoding/decoding rules, any one of the all eight addition rules and any one of the all eight subtraction rules randomly for each pixel under the control pseudorandom sequences generated through the conservative chaotic standard map.

**2.2 The generation of pseudorandom sequences based on conservative chaos and their testing:**

In the proposed image encryption algorithm, the conservative chaotic standard map is used in a novel way for the generation of pseudo-random number sequences which drives the entire process of DNA encryption. The following form of the 2D conservative chaotic standard map is used for this purpose.

$$X_{n+1} = f(X_n, Y_n) = (X_n + K \sin Y_n) \bmod 2\pi,$$

$$Y_{n+1} = g(X_n, Y_n) = (X_{n+1} + Y_n) \bmod 2\pi,$$

Here $X$ and $Y$ are the state variables and $K$ is the parameter. The chaotic region in the phase space increases with the increase in parameter K and chaos becomes completely global for K>18 and the chaotic orbit visits the entire phase space ergodically [57].

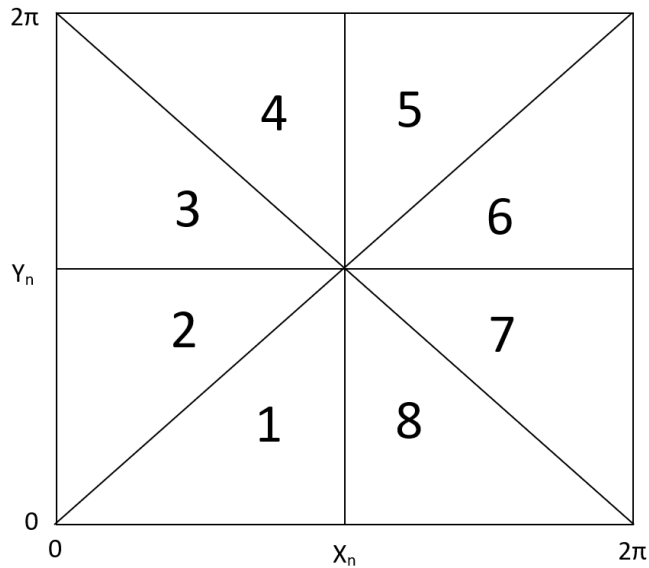

**Figure 1:** Divisions of the phase space of the conservative chaotic standard map

The iterates of the abovementioned map are used for the generation of pseudo-random number sequences. For this purpose, we divide the entire phase space $(0 < X < 2\pi, 0 < Y < 2\pi)$ of the conservative chaotic map into eight equal parts as depicted in Figure 1 and assign numbers 1 to 8 to these parts. After each iteration, we observe the pair of values of X and Y, depending on which region of the phase space this belongs to, we record the corresponding region number in the sequence. In this way, we generate a sequence comprising numbers 1 to 8 of the desired length using the iterates of the above-mentioned map.

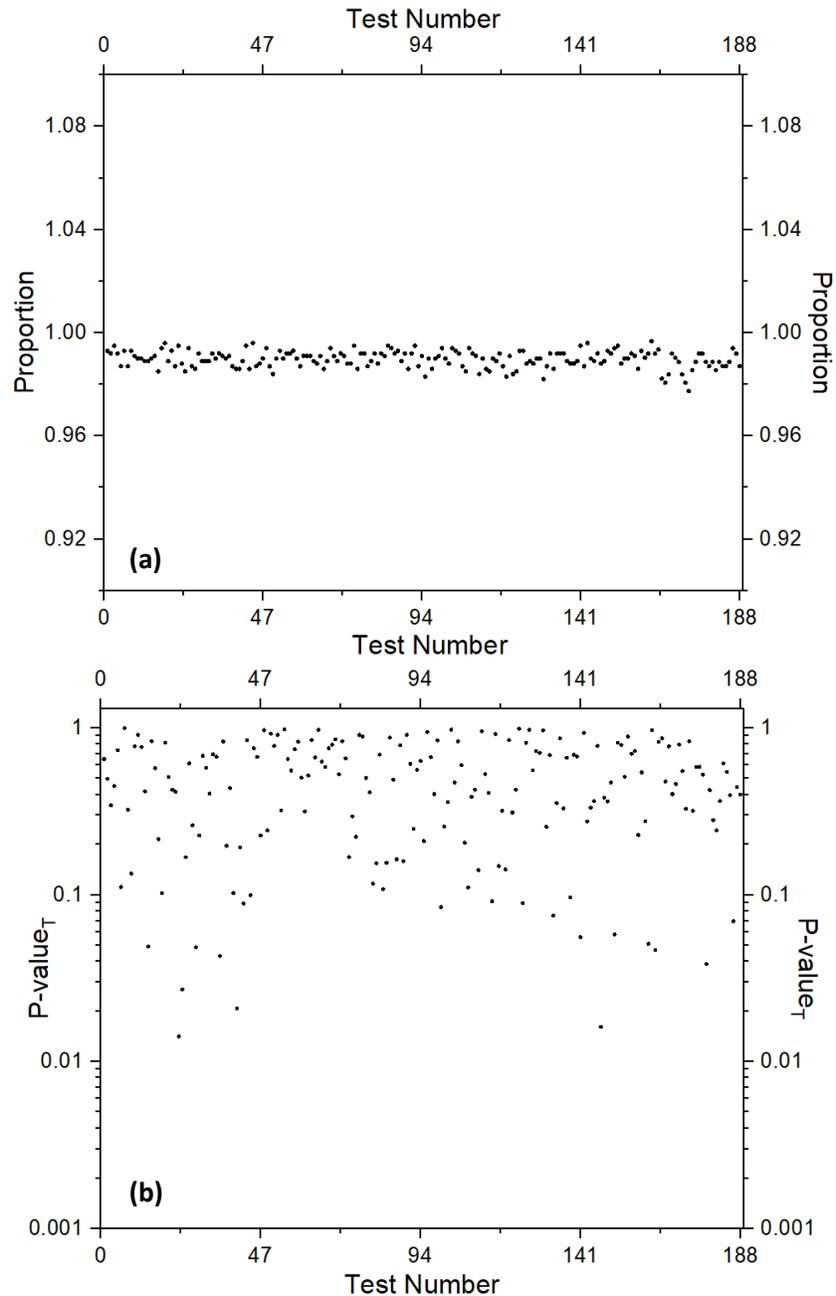

**Figure 2:** Testing of pseudorandom sequences using NIST test suite: (a) Proportions and (b) $p-\text{value}_T$

We have also tested the randomness of pseudorandom sequences generated in the above-mentioned manner using the NIST test suite [59]. For the testing purpose, we have generated 1000 sequences of $10^6$ bits each starting with random choices of initial conditions and parameter values of the conservative chaotic standard map and run the entire test suite comprising 15 different parametric and non-parametric tests (in total there are 188 tests which include all variants of different tests in the suite). For each sequence, using each test statistic, a p-value is generated. If the p-value is greater than 0.01 (determined by the chosen significance level) then the test is labelled as passed. A certain number of sequences, out of the total tested, are expected to fail the test depending on the level of significance chosen. The NIST test suite also predicts how many sequences out of the total sequences have passed the test, it is defined as the proportion (no. of sequences passing the test/total no. of sequences) of the sequences passing the test. For a significance level of 0.01, the allowed range of proportions is [ 0.9833245, 0.9966745]. In Figure 2(a). we have depicted the proportions for all 188 different tests for the testing set of 1000 sequences. It is observed that the proportion for all the tests falls within the allowed range of proportions. To check the uniformity of the distribution of all p-values (1000 in number) for a particular test, we obtain a $p-value_T$ using $\chi^2$ test (i.e. p-value of the p-values). If the $p-value_T$ is greater than 0.0001, it is declared that the sequences have uniformly distributed p-values for that test and the test is termed as passed. We have depicted the $p-value_T$ for the distribution of p-values for all 188 statistical tests in Figure 2(b) which indicates that the uniformity is observed for all the tests included in the NIST test suite. For more details on various test statistics and testing procedures, readers are referred to [59]. With these testing results, we may conclude that the pseudorandom sequences generated in the above manner are cryptographically secure and hence these sequences may be used in any encryption algorithm.

**2.3 The encryption process:**

The DNA operations are capable of shuffling as well as altering the pixel values therefore if implemented in a specific and strategic manner these operations may produce the desired permutation-substitution effect as recommended by Shanon [1]. For this purpose, in the proposed image encryption algorithm, we use a conservative chaos-driven dynamic DNA coding procedure. We use DNA encoding, addition/subtraction and decoding of pixels in the encryption (the procedure of DNA addition is replaced with the DNA subtraction in the decryption process). We use all eight possible DNA encoding rules and corresponding addition, subtraction and decoding rules in the proposed image encryption algorithm which are dynamically chosen for each pixel at various stages of encryption. All these processes are executed pixel-wise and the DNA encoding, addition, and decoding rules for each pixel are selected randomly with the help of pseudorandom number sequences generated through the conservative chaotic standard map. While executing the two-step DNA addition, we bring in the ciphertext dependence through the feedback mechanism wherein the last cipher pixel is also used in the second step of DNA addition (see Step 12 below). Before executing the DNA operations as explained above, we also use the conservative chaotic standard map to generate a dynamic one-time pixel (DOTP) value for the encryption of each pixel of the plain image. The DOTP is generated in such a way that it also possesses the sensitive dependence on all the plain image pixels ahead of the pixel being encrypted (feed-forward mechanism) as well as on all the cipher image pixels generated before the encryption of that pixel (feedback mechanism) (see Steps 7 and 8 below). Before introducing this plain image and cipher image sensitivity, we also use DNA encoding of the DOTP using the randomly selected DNA encoding rule controlled by a pseudo-random sequence generated through the conservative chaotic standard map. Below we give the process flow and finer details of the entire image encryption algorithm.

The proposed image encryption algorithm has been explained for a grey image of size $H \times W$ as the plain image. However, it can be easily extended to RGB images by converting/reshaping the 3D RGB pixels matrix to a 2D matrix and considering it as the input to this algorithm. The other way is to encrypt all three layers separately. The secret key in the proposed image encryption algorithm is a set of one integer and seven floating-point numbers. The two floating-point numbers $(X_0, Y_0) \in (0, 2\pi)$ serve as the initial conditions for the chaotic conservative standard map, the remaining five floating-point numbers $(K, K_1, K_2, K_3, K_4) > 18.0$ serve as the parameter value for the conservative chaotic standard map at various stages of the algorithm and an integer $0 < N < 1000$ serves as the number of iterations to skip before using the map for the encryption purpose. The entire process of proposed image encryption can be divided into two parts:

## Part-I

This part of the encryption process requires the secret key and the information on the size of the plain image. If in certain applications e.g., online streaming of videos in TV broadcasting through viewing cards where the size of images and secret keys are fixed, this part of the encryption process may be pre-computed and stored to speed up the encryption. This part of the encryption process can be identified with the red dotted block in the block diagram of the encryption process in Figure 3.

1. The conservative chaotic standard map is iterated N number of times with the initial conditions $(X_0, Y_0)$ and parameter $K$ specified in the secret key. The iterates are thrown out and only the values $(X_N, Y_N)$ are stored for further use.
2. The conservative chaotic standard map is iterated HXW number of times with the initial conditions $(X_N, Y_N)$ and parameter $K$, all the iterates X and Y are used in the following way to generate the DOTP1 and DOTP2.
$DOTP1(k) = \left\lfloor \frac{X_{N+k}}{2\pi} \times 256 \right\rfloor, k = 1, 2, \ldots, H \times W,$
$DOTP2(k) = \left\lfloor \frac{X_{N+k}}{2\pi} \times 256 \right\rfloor, k = 1, 2, \ldots, H \times W$
3. The conservative chaotic standard map is iterated HXW number of times with the initial conditions $(X_{N+HW}, Y_{N+HW})$ and parameter $K_1$. All the iterates X and Y are used to generate a pseudo-random number sequence $RSQ1_i$ ($i = 1$ to $H \times W$) having integers 1 to 8 using the procedure explained in subsection 2.2.
4. Repeat Step 3 with the initial condition $(X_{N+2HW}, Y_{N+2HW})$ and parameter $K_2$ to generate a pseudo-random number sequence $RSQ2_i$ ($i = 1$ to $H \times W$) having integers 1 to 8.
5. Repeat Step 3 with the initial condition $(X_{N+3HW}, Y_{N+3HW})$ and parameter $K_3$ to generate a pseudo-random number sequence $RSQ3_i$ ($i = 1$ to $H \times W$) having integers 1 to 8.
6. Repeat Step 3 with the initial condition $(X_{N+4HW}, Y_{N+4HW})$ and parameter $K_4$ to generate a pseudo-random number sequence $RSQ4_i$ ($i = 1$ to $H \times W$) having integers 1 to 8.

## Part-II

This part of encryption is executed pixel-wise and therefore repeated $H \times W$ times. Here we are explaining the process for the $i^{th}$ pixel of the plain image. This part of the encryption process may be identified with the green dotted block in the block diagram of the encryption process in Figure 3.

7. Compute two terms plain image dependent term (PIDT) and cipher image dependent term (CIDT) in the following way.

$$PIDT(i) = mod(\sum_{k=i+1}^{H \times W} PI_k, 256)$$

$$CIDT(i) = mod(\sum_{k=1}^{i-1} CI_k, 256)$$

Here $PI_k$ and $CI_k$ are $k^{th}$ plain image and cipher image pixels. For the first pixel of the plain image the value $CIDT$ will be zero.

8. Compute and modify the DOTP for the encryption of $i^{th}$ pixel

$$DOTP(i) = \left((DOTP1(i) \oplus DOTP2(i)) \oplus PIDT(i)\right) \oplus CIDT(i),$$

here $\oplus$ is the XOR operation. This step of encryption brings in extreme sensitivity to the plain image and cipher image too and makes it very robust against known plaintext and chosen-ciphertext attacks.

This step onwards the role of conservative chaos-driven DNA encoding, decoding and addition starts.

9. Encode the $DOTP(i)$ in DNA sequence using the DNA $RSQ1_i^{th}$ Encoding Rule. Here first we convert the $DOTP(i)$ value in the 8-bit binary form.
10. Encode the $i^{th}$ pixel of the plain image (first converted to 8-bit binary form) i.e. $PI_i$ in DNA sequence using the DNA $RSQ2_i^{th}$ Encoding Rule.
11. Add the DNA sequences of $DOTP(i)$ and $PI_i$ using the DNA $RSQ3_i^{th}$ Addition Rule.
12. Now we generate the DNA sequence of the $i^{th}$ pixel of cipher image in the following way: The resultant DNA sequence from Step 11 is added to the DNA sequence of the $(i-1)^{th}$ pixel of cipher image using the DNA $RSQ3_i^{th}$ Addition Rule. For the encryption process of the first plain image pixel, the DNA sequence of the cipher image pixel to be added here is fixed to 'ATCG' as the $0^{th}$ cipher image pixel does not exist.
13. DNA sequence of the $i^{th}$ pixel of cipher image (generated in Step 12) is converted to the 8-bit binary form using the DNA $RSQ4_i^{th}$ Decoding Rule.
    The decimal equivalent of this 8-bit is finally considered as the intensity of the $i^{th}$ pixel of the cipher image $CI_i$.

The entire Part-II of the proposed encryption (except Steps 7 and 8) is based on the dynamic lookup table operations (DNA encoding, DNA addition, DNA subtraction, DNA decoding tables) controlled by the pseudo-random sequences generated in the Part-I, therefore, almost negligible arithmetic operations are involved in Part-II of the encryption and hence can be executed very fast.

In the next section, we analyze the performance and security of the proposed image encryption algorithm through various statistical and perceptual quality analyses.

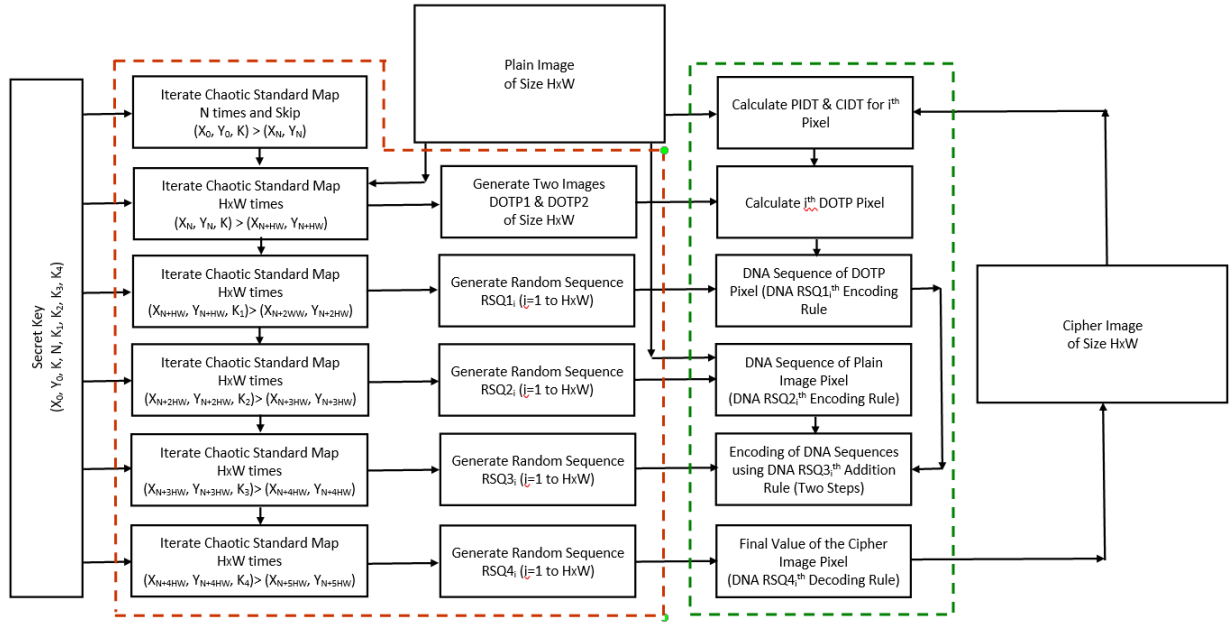

**Figure 3:** Block diagram of the proposed image encryption process

## 3. Performance and Security Analysis:

For the performance and security analysis of the proposed image encryption algorithm, we have used five different images *Lena*, *Baboon*, *Peppers*, *AllBlack* (all pixel values are '0') and *AllWhite* (all pixel values are '255') each of size 200 X 200. For the encryption of these images, we have considered five different randomly chosen secret key combinations which are depicted in Figure 4 along with their corresponding cipher images produced using the proposed image encryption algorithm.

### 3.1 Histogram Analysis:

For an ideal cipher image, the histogram must be uniform i.e. the number of pixels corresponding to all intensity levels should be equal irrespective of the content of the plain image as well as of the secret key. The histograms for all five pairs of plain and cipher images are also shown in Figure 5. From the visual inspection of cipher images and their histograms, we may easily infer that the histograms are uniform.

However, for quantitative confirmation, we have also computed the statistical measures like chi-square distribution, histogram variance, deviation from ideality, maximum deviation and irregular deviation for the histograms which mainly confirms the uniformity of the cipher image histograms and also predicts the amount of deviation between the histograms of plain and cipher images. The details of these measures and the results of our analysis are described in the following subsections.

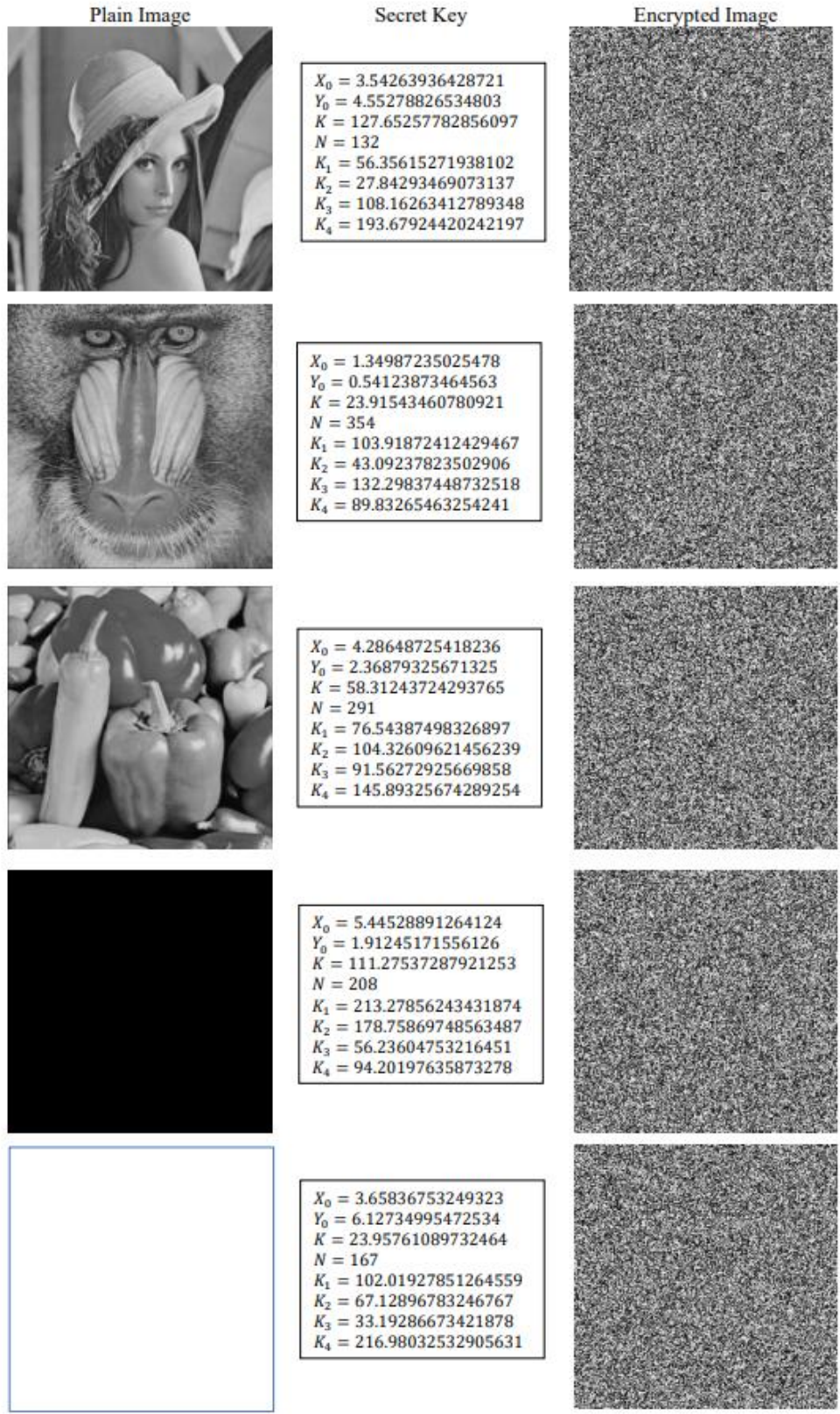

**Figure 4:** Plain images, corresponding secret keys and cipher images

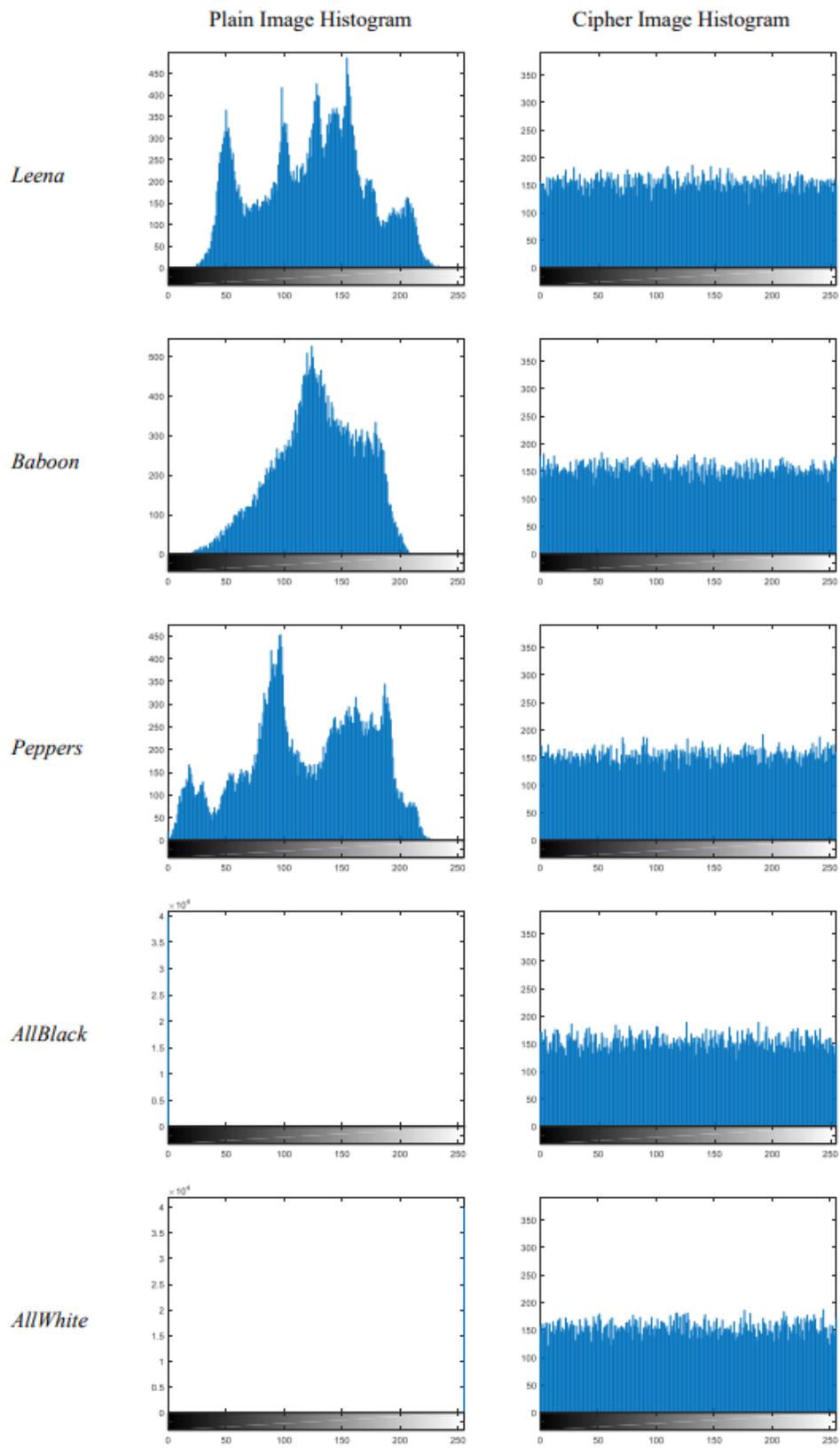

**Figure 5:** Histograms of Plain images and corresponding cipher images

## (i) Chi-square and histogram variance:

To understand the pixel distribution quantitively, we compute the $\chi^2$ from the histograms of the plain and cipher images using the following statistical formula.

$$\chi^2(I) = \sum_{i=1}^{256} \frac{(f_i - f_0)^2}{f_0},$$

here $f_0 = \frac{total\ number\ of\ pixels}{number\ of\ intensity\ levels\ i.e. 256}$ and $f_i$ is the total number of pixels at the $i^{th}$ intensity level.

For perfect uniform distribution, the value of $\chi^2$ is zero and a standard value of the $\chi^2$ for typically acceptable random cipher image (significance level 0.05) is around 293. We have displayed the results of $\chi^2$ for all five pairs of plain and cipher images in Table 5. We observe that the $\chi^2$ values for plain images are very high and it is lower than the acceptable standard value for the cipher images produced using the proposed image encryption algorithm.

Table 5: $\chi^2$ and Histogram Variance for the plain and cipher images

|  |  | Lena | Baboon | Peppers | All Black | All White |
|---|---|---|---|---|---|---|
| $\chi^2$ | Plain Image | 2.5400e+04 | 3.8312e+04 | 1.9280e+04 | 10200000 | 10200000 |
|  | Cipher Image | 241.4336 | 220.8128 | 229.7856 | 291.7888 | 277.7984 |
| HistVar | Plain Image | 1.5503e+04 | 2.3384e+04 | 1.1768e+04 | 6.2256e+06 | 6.2256e+06 |
|  | Cipher Image | 147.3594 | 134.7734 | 140.2500 | 178.0938 | 169.5547 |

We have also calculated the histogram variances (*HistVar*) for the plain and cipher images using the following statistics:

$$HistVar(I) = \frac{1}{N^2} \sum_{i=1}^{N} \sum_{j=1}^{N} \frac{1}{2}(f_i - f_j)^2,$$

here $f_i$ and $f_j$ are the total number of pixels at the $i^{th}$ and $j^{th}$ intensity levels respectively, and $N$ is the total number of intensity levels i.e. 256. The results of our computation for all the five pairs of plain and cipher images are shown in Table 5, the results clearly show that the histogram variances are very high for the plain images and very low for the cipher images (almost 1% of the histogram variance of plain images).

## (ii) Deviation from ideality (DI), Maximum Deviation (MD) and Irregular Deviation (ID)Chi-square and histogram variance:

Another way to measure the uniformity of the histograms of cipher images is using a metric Deviation from Ideality (DI). It measures the deviation of the histogram of the cipher image from the ideal uniform histogram. The DI metric is calculated in the following way:

$$DI = \frac{1}{H \times W} \left( \sum_{i=0}^{255} (f_i(C) - f_0) \right)$$

here $f_0 = \frac{\text{total number of pixels}}{\text{number of intensity levels i.e.256}}$ and $f_i(C)$ is the histogram of $i^{th}$ level (total number of pixels at the $i^{th}$ intensity level ) in the encrypted image.

The histogram of the cipher image is nearly uniform if the value of the DI metric is nearly zero or very low. For a completely uniform/ideal histogram, the value of DI is zero. In Table 6, we have shown the values of DI metric for all five cipher images corresponding to the five plain images used in the analysis. We observe that the DI values are nearly zero or negligible hence the histograms of cipher images are almost uniform.

Similarly, another metric referred as Maximum Deviation (MD) measures the deviation of the histogram of the cipher image from the histogram of the plain image. The computation of MD can be done using the following statistics:

$$MD = \frac{1}{H \times W}\left(\frac{d_0 + d_{255}}{2} + \sum_{i=1}^{254} d_i\right),$$

here $d_i$ is the absolute difference between the histograms corresponding to the $i^{th}$ level of plain and encrypted images. The higher the value of MD, the larger the deviation between the histograms of the plain and cipher images. We have shown the values of metric MD also in Table 6 for all five pairs of plain and cipher images in the present analysis. The results clearly show that the histograms of plain and cipher images are significantly different.

Sometimes, the Maximum Deviation does not provide the correct information about the deviation between the histograms therefore may mislead the interpretation. To overcome this, another metric Irregular Deviation (ID) is also used which measures the deviation of the difference of the histograms between plain and cipher images with the mean of the difference of the histograms and high value of ID signifies a better encryption algorithm. The metric ID is calculated using the following formula/statistics:

$$ID = \frac{1}{H \times W}\left(\sum_{i=0}^{255}|d_i - M_d|\right),$$

here $d_i$ is the absolute difference between the histograms corresponding to the $i^{th}$ level of plain and encrypted images and $M_d$ is the mean of the difference of histograms. The higher values signify the larger deviation. The results for the metric ID for all five pairs of plain and cipher images are given in Table 6. The results indicate that the plain and cipher images are significantly different in terms of statistical deviations.

**Table 6:** Deviation of cipher images from ideality, maximum and irregular deviations between plain and cipher images

|    | Lena   | Baboon | Peppers | All Black | All White |
|----|--------|--------|---------|-----------|-----------|
| DI | 0.0615 | 0.0597 | 0.0595  | 0.0705    | 0.0671    |
| MD | 0.6507 | 0.8778 | 0.5731  | 1.4916    | 1.4916    |
| ID | 0.6937 | 0.7622 | 0.6459  | 0.0705    | 0.0671    |

All the above results of the histogram analysis confirm the desired level of uniformity of pixel distribution in the cipher images and remove the possibility of implementing statistical attacks based on histogram analysis.

## 3.2 DNA Sequence-based analysis:

### (i) Hamming Distance (HD)

Hamming distance is used to compare two character/symbol strings of equal length and is defined as the number of positions where two corresponding symbols/characters are different in two strings. For two image DNA sequences $I_1$ and $I_2$ the Hamming distance (HD) is defined as:

$$HD(I_1, I_2) = \sum_{i=1}^{H \times W \times 4} d(I_{1i}, I_{2i}) \text{ where } d(I_{1i}, I_{2i}) = \begin{cases} 0 \text{ if } I_{1i} = I_{2i} \\ 1 \text{ if } I_{1i} \neq I_{2i} \end{cases},$$

$I_{1i}$ and $I_{2i}$ are the $i^{th}$ symbol /base in the DNA sequence of the images $I_1$ and $I_2$ respectively, $H$ and $W$ are the height and width of images $I_1$ and $I_2$.

The higher the value of Hamming distance, the more dissimilar the strings are. In our analysis, we are comparing the DNA sequences of plain images and cipher images. Considering the images of 200 X 200 size used in the test run, there are total 40000 pixels in each image and hence 160000 bases in their corresponding DNA sequences. The results of our computation for the Hamming distance for all five pairs of plain and cipher images are recorded in Table 7. It is observed that in all the cases the Hamming distance is nearly 120000 irrespective of the content of the plain image as well as the secret key used for the encryption. It signifies that 75% of the bases in the DNA sequences of plain and cipher images are different.

**Table 7:** Hamming distance (HD) between plain and cipher images

| Lena | Baboon | Peppers | All Black | All White |
|---|---|---|---|---|
| 120078 | 119907 | 119749 | 120023 | 120126 |

### (ii) Base Ratio (BR):

The base Ratio (BR) of a particular base in a DNA sequence is the percentage of the occurrence of that base in the sequence. For an image of height H and width W, the base ratio can be calculated in the following way

$$BR(S) = \frac{count(S)}{H \times W \times 4} \times 100\%,$$

where S is one of the symbol/base (out of A, T, C and G) in the DNA sequence of an Image. As every pixel in the image is represented by four DNA bases therefore the total symbols in the DNA sequence of an image shall be four times the number of image pixels. We have shown the results of the base ratio for all four bases (A, T, C and G) for all five pairs of plain and cipher images in Table 8. We observe that in all cases the occurrence of DNA bases is uniform (nearly 25%) in both the plain and cipher images. It is interesting to note here that in the proposed image encryption algorithm chaos-based randomized encoding of the plain image is used as the first step of encryption and the base ratio results show that this first step of encoding itself brings in so much uniformity in the encryption process. This is testimony to the fact that the proposed image encryption method is robust against any sort of statistical attack on the DNA bases.

Table 8: Base Ratio (BR) for various plain and corresponding cipher images

|  | DNA Base | Lena | Baboon | Peppers | All Black | All White |
|---|---|---|---|---|---|---|
| Plain Image | A | 24.9319 | 24.9581 | 25.0738 | 25.3900 | 25.1000 |
|  | T | 24.9744 | 24.8744 | 24.9919 | 24.4825 | 25.0875 |
|  | C | 24.9163 | 25.0794 | 24.8881 | 24.6575 | 24.7450 |
|  | G | 25.1775 | 25.0881 | 25.0463 | 25.4700 | 25.0675 |
| Cipher Image | A | 25.0894 | 24.9662 | 24.8025 | 25.2475 | 24.9975 |
|  | T | 25.1625 | 25.1406 | 24.8850 | 24.8381 | 24.8400 |
|  | C | 24.7319 | 24.9537 | 25.1850 | 24.8006 | 25.1563 |
|  | G | 25.0162 | 24.9394 | 25.1275 | 25.1138 | 25.0063 |

### 3.3 Fixed-Point Ratio (FPR):

A particular pixel in an image is identified as the fixed-point if it does not change its grey value after the entire encryption process. The fixed-point ratio is the percentage number of such fixed points which exist in an image after the encryption. For a pair of plain and cipher images, the fixed-point ratio (FPR) is calculated in the following way:

$$FP(P,C) = \frac{\sum_{i=1}^{W \times H} f(P_i, C_i)}{H \times W} \times 100\%, \text{ with } f(P_i, C_i) = \begin{cases} 1 \text{ if } P_i = C_i \\ 0 \text{ if } P_i \neq C_i \end{cases},$$

where $P_i$ and $C_i$ are the $i^{th}$ pixel values in the images $P$ and $C$ respectively, $H$ and $W$ are the height and width of images $P$ and $C$.

For all five pairs of plain and cipher images, we have summarized the values of the FPR metric in Table 9, these results clearly show that the percentage of pixels which do not change after encryption through the proposed method is below 0.5% and thus it signifies the existence of effective substitution and diffusion in the proposed image encryption algorithm.

Table 9: Fixed Point Ratio (FPR) for various pairs of plain and cipher images

| Lena | Baboon | Peppers | All Black | All White |
|---|---|---|---|---|
| 0.4000 | 0.3550 | 0.4025 | 0.4300 | 0.4300 |

### 3.4 Correlation analysis:

In an image having definite visual content, the adjacent pixels are highly correlated and a weak encryption process does not completely remove such correlations. In addition to this, cryptanalysts sometimes use pairs of plain and corresponding cipher images to identify some meaningful relationship between the plain and cipher images by analyzing the correlation between the pairs of plain and cipher images. An ideal cipher should produce cipher images possessing almost zero correlation with the plain images.

To analyze the above mention types of correlations, we have computed the correlation coefficients for all horizontally and vertically adjacent pixel pairs in all plain and their corresponding cipher images using the following expressions:

$$C = \frac{\frac{1}{N}\sum_{i=1}^{N}(x_i-\bar{x})(y_i-\bar{y})}{\sqrt{\left(\frac{1}{N}\sum_{i=1}^{N}(x_i-\bar{x})^2\right)\left(\frac{1}{N}\sum_{i=1}^{N}(y_i-\bar{y})^2\right)}} \text{ with } \bar{x} = \frac{1}{N}\sum_{i=1}^{N}x_i \text{ and } \bar{y} = \frac{1}{N}\sum_{i=1}^{N}y_i,$$

here $x_i$ and $y_i$ form $i^{th}$ pair of horizontally/vertically adjacent pixels and $N$ is the total number of pairs of horizontally/vertically adjacent pixels.

In general, for plain images having definite visual content, the correlation coefficients are very high and ideally, for cipher images, these correlation coefficients should be negligible or zero. The results of such horizontal and vertical correlation coefficients have been given in Table 10. The results indicate that there is no correlation between plain and cipher image pixels thereby eliminating the possibility of implementing any statistical attack based on the correlation.

**Table 10:** Correlation between horizontally & vertically adjacent pixels in plain & cipher images and 2D Correlation between pairs of plain & cipher images

|  |  | Lena | Baboon | Peppers | All Black | All White |
|---|---|---|---|---|---|---|
| Horizontal Adjacent Pixels | Plain Image | 0.9322 | 0.8670 | 0.9544 | 0.5774 | 0.5774 |
|  | Cipher Image | 3.8945e-04 | -0.0019 | -0.0067 | -0.0030 | -0.0078 |
| Vertical Adjacent Pixels | Plain Image | 0.9684 | 0.8315 | 0.9646 | 0.5774 | 0.5774 |
|  | Cipher Image | -0.0083 | -0.0028 | 0.0054 | -0.0034 | -0.0028 |
| 2D Correlation between Plain and Cipher Image |  | -0.0016 | 0.0019 | 0.0013 | 7.0171e-04 | 0.0076 |

We have also computed the 2D correlation coefficient between the plain and corresponding cipher image using the following statistics:

$$C_{PC} = \frac{\frac{1}{H\times W}\sum_{i=1}^{H}\sum_{j=1}^{W}(P_{ij}-\bar{P})(C_{ij}-\bar{C})}{\sqrt{\left(\frac{1}{H\times W}\sum_{i=1}^{H}\sum_{j=1}^{W}(P_{ij}-\bar{P})^2\right)\left(\frac{1}{H\times W}\sum_{i=1}^{H}\sum_{j=1}^{W}(C_{ij}-\bar{C})^2\right)}} \text{ with } \bar{P} = \frac{1}{H\times W}\sum_{i=1}^{H}\sum_{j=1}^{W}P_{ij} \text{ and } \bar{C} = \frac{1}{H\times W}\sum_{i=1}^{H}\sum_{j=1}^{W}C_{ij},$$

here $P$ and $C$ are the plain and encrypted images respectively.

The result of such 2D correlation coefficients for all five pairs of plain and cipher images are given in Table 10, which clearly shows that the pairs of plain and cipher images do not possess any correlation, therefore, removing the possibility of implementing statistical attacks based on correlation.

### 3.5 Information entropy analysis:

Information entropy (also referred to as Shanon entropy or global information entropy) is a measure of uncertainty associated with a random image or it may be considered as the measure of disorder. It quantifies the amount of information contained in the image (in bits) per pixel. It can also be interpreted as the minimum number of bits per pixel necessary to communicate it correctly. It is also a statistical measure of randomness in the image. The information entropy for a greyscale image may be computed in the following way:

$H(I) = \sum_{i=1}^{256} P(I_i) \log_2 \frac{1}{P(I_i)}$ (bits per pixel), where $P(I_i)$ is the probability of occurrence of the pixel value $I_i$ in the image $I$. For a flat image, the information entropy is zero and for an image whose pixel distribution is perfectly uniform (i.e., $P(I_i) = 1/256$) the information entropy is 8-bits per pixel. The results of global information entropy for the plain and cipher images are given in Table 11. It is observed that the global information entropy for the encrypted images is very close to the maximum possible value i.e 8-bits.

**Table 11:** Global and Local information Entropy

|  |  | Block Size | Lena | Baboon | Peppers | All Black | All White |
|---|---|---|---|---|---|---|---|
| Global Information Entropy | Plain Image | 200 X 200 | 7.4351 | 7.1938 | 7.5820 | 0 | 0 |
|  | Cipher Image |  | 7.9956 | 7.9960 | 7.9959 | 7.9947 | 7.9950 |
| Local Information Entropy | Plain Image | 50 X 50 | 6.6886 | 6.8043 | 6.9406 | 0 | 0 |
|  | Cipher Image |  | 7.9253 | 7.9240 | 7.9259 | 7.9266 | 7.9213 |
|  | Plain Image | 40X 40 | 6.5113 | 6.7134 | 6.7164 | 0 | 0 |
|  | Cipher Image |  | 7.8804 | 7.8804 | 7.8841 | 7.8810 | 7.8798 |
|  | Plain Image | 25 X 25 | 5.7811 | 6.4054 | 6.2370 | 0 | 0 |
|  | Cipher Image |  | 7.4703 | 7.6699 | 7.6730 | 7.6742 | 7.6671 |

The Shanon entropy or global information entropy measure may possess some weaknesses such as inaccuracy, inconsistency, and low efficiency in certain cases and to overcome such weakness, a new variant named local information entropy is suggested which is the mean entropy of several or all non-overlapping image blocks that are randomly selected from image. For an image $I$ divided into $k$ number of non-overlapping blocks $I_i$ (i = 1 to k), the local information entropy may be computed in the following manner:

$H(I)_{local} = \frac{1}{k} \sum_{i=1}^{k} H(I_i)$,

here $H(I_i)$ is the Shanon entropy of the $i^{th}$ block $I_i$ and $k$ is the total number of non-overlapping blocks of the image $I$. The result of local information entropy for the plain and cipher images corresponding to three different block sizes (25X25, 40 X 40, 50 X 50) are given in Table 11. We observe that the local information entropy of cipher images is also close to the global information entropy and well above the standard values of local information entropy of random images.

### 3.6 Perceptual quality analysis:

One of the major objectives of the image encryption algorithm is to secure the content by making the unintelligible and obfuscating the visual data to appear random. It can be observed from the Figure 4, that after the encryption the images look completely random with no visual patterns/content. In addition to the visual inspection of the encrypted images, quantitative perceptual quality analysis is also done for the image encryption processes to observe how much quality degradation is introduced (of course recoverable at the decryption) by the encryption algorithm so that the information becomes completely unintelligible and appear garbage. For an encryption algorithm, it is expected that encrypted images have low perceptual quality with reference to the plain image and it is measured with metrics such as mean absolute error (MAE), Mean square error (MSE), peak-signal-to-noise ratio (PSNR), spectral distortion (SD), structural similarity index measure (SSIM) and feature similarity index measure (FSIM).

The MAE, MSE and PSNR are used to quantify the image fidelity or spatial dissimilarities between two images. Although these metrics do not include the characteristics of image signal and the human vision system (HSV), they are widely used as the first full-reference measures. In encryption, it is expected to have large values of MAE and MSE and low values of PSNR (<<28) which convey the higher amount of average dissimilarity in the pixel values between the plain and cipher images. These metrics may be computed in the following way:

$$MAE = \frac{1}{H \times W} \sum_{i=1}^{H} \sum_{j=1}^{W} |P_{ij} - C_{ij}|,$$

$$MSE = \frac{1}{H \times W} \sum_{i=1}^{H} \sum_{j=1}^{W} |P_{ij} - C_{ij}|^2 \text{ and}$$

$$PSNR = 10 \log_{10} \frac{(Max(f))^2}{\frac{1}{H \times W} \sum_{i=1}^{H} \sum_{j=1}^{W} |P_{ij} - C_{ij}|^2},$$

here $P_{ij}$ and $C_{ij}$ are the $ij^{th}$ pixel values of the plain and encrypted images respectively and $Max(f)$ is the highest intensity level i.e., 255.

To measure the spectral dissimilarity between the plain and encrypted images, the spectral distortion (SD) measure is used. It is computed using the following expression: $SD = \frac{1}{H \times W} \sum_{u=1}^{H} \sum_{v=1}^{W} |F_p(u,v) - F_c(u,v)|$, here $F_p$ and $F_c$ are the discrete Fourier tramsform of the plain and encrypted images respectively.

Another metric structural similarity index measure (SSIM) takes into consideration the human vision system (HSV) and compares the images with respect to luminance, contrasts and structural features [60] For perfectly identical images SSIM is '1' and very low for dissimilar images with respect to the above features. It can be computed using the following statistics: $SSIM(a,b) = \frac{(2\mu_a\mu_b + C_1)(2\sigma_{ab} + C_2)}{(\mu_a^2 + \mu_b^2 + C_1)(\sigma_a^2 + \sigma_b^2 + C_2)}$ where $\mu_a$ is the average of all pixels of image a, $\mu_b$ is the average of all pixels of image b, $\sigma_a^2$ is the variance of the pixel values of image a, $\sigma_b^2$ is the variance of the pixel values of image b and $\sigma_{ab}$ is the covariance of pixels of images a and b.

Another comparatively new perceptual image quality measure: the feature similarity index measure (FSIM) takes into consideration the phase congruency (PC) and gradient magnitude (GM) as two complementary feature measures to characterize the image local quality. FSIM metric is computed for two images in the following way [61]:

$FSIM(a,b) = \frac{2f_a f_b + c}{f_a^2 + f_b^2 + c}$, here $f$ indicates one of the features (PC or GM). FSIM is computed individually for the PC and GM and then multiplied to obtain the final FSIM. For perfectly similar images FSIM is '1' and low for the dissimilar images with respect to PC and GM.

We have computed all six above explained measures for all the five pairs of plain and cipher images, to observe the perceptual quality of the encrypted images produced using the proposed image encryption algorithm and the results have been summarized in Table 12. It can be easily observed from the results that as desired for an encrypted image the MAE, MSE are very high, PSNR is very low, SD is very high and SSIM and FSIM are small which confirms the very low perceptual quality of encrypted images.

Table 12: Perceptual quality metrics

|      | Lena      | Baboon    | Peppers   | All Black  | All White  |
|------|-----------|-----------|-----------|------------|------------|
| MAE  | 72.5634   | 69.4042   | 75.1730   | 127.3586   | 127.5093   |
| MSE  | 7.6628e+03 | 6.8391e+03 | 8.3134e+03 | 2.1668e+04 | 2.1716e+04 |
| PSNR | 9.2869    | 9.7808    | 8.9330    | 4.7726     | 4.7629     |
| SD   | 1.3971e+04 | 1.3885e+04 | 1.4271e+04 | 1.3225e+04 | 1.3211e+04 |
| SSIM | 0.0125    | 0.0092    | 0.0111    | 4.7659e-06 | 0.0090     |
| FSIM | 0.3618    | 0.4806    | 0.3695    | 0.0461     | 0.0712     |

### 3.7 Plaintext sensitivity analysis (differential analysis):

To resist the differential analysis in which the attacker may analyze the relationship between the plaintext and ciphertext by making minor changes in the plaintext and observing the effects in the ciphertext to discover the secret key. To check the robustness of the encryption algorithm against such differential analysis, we quantify the plaintext sensitivity of the encryption algorithm using two metrics Net Pixel Change Rate (NPCR) which measures the percentage number of pixels in the encrypted image which change their values after making an infinitesimal change in the plaintext and encrypted with the same secret key and Unified Average Change Intensity (UACI) which measures the net average change in the intensity of each pixel in the encrypted image after making an infinitesimal change in the plaintext and encrypted with the same secret key. The computation is done by comparing the two cipher images which are produced using the same secret key and their corresponding plaintexts are differing in only one-pixel value at any random location. Several random combinations of the secret key and locations of the pixel in plain image are considered one by one and average values of NPCR and UACI are computed. Following mathematical formulae are used for the computation of the NPCR and UACI.

$NPCR = \left(\frac{1}{H \times W} \sum_{i=1}^{H} \sum_{j=1}^{W} D_{ij}\right) \times 100\%$, with $D_{ij} = \begin{cases} 0 \text{ if } C_{ij} = \bar{C}_{ij} \\ 1 \text{ if } C_{ij} \neq \bar{C}_{ij} \end{cases}$,

$UACI = \left(\frac{1}{H \times W} \sum_{i=1}^{H} \sum_{j=1}^{W} \frac{|C_{ij} - \bar{C}_{ij}|}{255}\right) \times 100\%$.

$C_{ij}$ and $\bar{C}_{ij}$ are two different cipher images produced with the same secret key and for slightly different plain images (only one pixel different). The standard values of NPCR and UACE for two random images are 99.6094 and 33.4635 respectively. The results of our computation for the NPCR and UACI for various plain images used in our analysis are summarized in Table 13. It can be observed that the NPCR and UACI for the proposed image encryption algorithm converge to the values for standard random images hence the two cipher images corresponding to two plain images having an infinitesimal difference are almost random therefore the proposed image encryption algorithm has required plaintext sensitivity to resist the differential attacks.

**Table 13:** NPCR and UACI in the proposed image encryption algorithm

|      | Lena    | Baboon  | Peppers | AllBlack | AllWhite |
|------|---------|---------|---------|----------|----------|
| NPCR | 99.6125 | 99.6825 | 99.6450 | 99.6225  | 99.6052  |
| UACI | 33.5115 | 33.4323 | 33.4561 | 33.5393  | 33.5060  |

### 3.8 Key sensitivity analysis:

In general, an ideal encryption algorithm should possess a complex and sensitive relationship between the secret key, plaintext and ciphertext. One of the ways to measure this sensitive behaviour is to observe the key sensitivity of the encryption algorithm. The key sensitivity may be measured in two ways: one at the encryption level and another at the decryption level.

To observe the key sensitivity at the encryption level, we encrypt the same plain image with two slightly different secret keys (differing by an infinitesimal change) and compare the two cipher images by computing KS1: the percentage of the total number of corresponding pixels which are different in both cipher images and KS2: the average change in the intensity of corresponding pixels in both the cipher images. It is done by using the following formulae

$$KS1 = \left(\frac{1}{H \times W} \sum_{i=1}^{H} \sum_{j=1}^{W} S_{ij}\right) \times 100\% \text{ with } S_{ij} = \begin{cases} 0 \text{ if } C_{ij}^A = C_{ij}^B \\ 1 \text{ if } C_{ij}^A \neq C_{ij}^B \end{cases}$$

$$KS2 = \left(\frac{1}{H \times W} \sum_{i=1}^{H} \sum_{j=1}^{W} \frac{|C_{ij}^A - C_{ij}^B|}{255}\right) \times 100\%.$$

Here $C_{ij}^A$ and $C_{ij}^B$ are two different cipher images corresponding to the same plain image produced with a minute change in one of the parts of the secret key.

In the proposed image encryption algorithm, there are eight parts of the secret key, seven of them are floating-point numbers and one is an integer. For computing the key sensitivity metrics, we make a change of $10^{-14}$ in only one of the parts of the secret key (if it is a floating-point number) or a change of 1(if it is an integer) and then compute KS1 and KS2 for the corresponding cipher images produced for the same plain image. The results of our computation for all five plain images are summarized in Table 14. The top row in the table indicates the part of the secret key which has been changed in the above-mentioned manner to compute KS1 and KS2. The results converge to the values for standard random images hence the two cipher images compared are perfectly random and therefore proposed image encryption algorithm possesses the extreme key sensitivity at the encryption level.

**Table 14:** Key sensitivity analysis results at the encryption level

|  |  | $X_0$ | $Y_0$ | $K$ | $N$ | $K_1$ | $K_2$ | $K_3$ | $K_4$ |
|---|---|---|---|---|---|---|---|---|---|
| Lena | KS1 | 99.6000 | 99.6275 | 99.6075 | 99.6075 | 99.6450 | 99.5450 | 99.6175 | 99.5375 |
|  | KS2 | 33.5236 | 33.6311 | 33.4323 | 33.3443 | 33.3465 | 33.3965 | 33.3671 | 33.4881 |
| Baboon | KS1 | 99.6075 | 99.5950 | 99.6075 | 99.5775 | 99.6250 | 99.5750 | 99.6400 | 99.5725 |
|  | KS2 | 33.1785 | 33.4433 | 33.3673 | 33.3415 | 33.4220 | 33.5358 | 33.5419 | 33.4749 |
| Peppers | KS1 | 99.6200 | 99.6225 | 99.6300 | 99.6025 | 99.6400 | 98.7575 | 98.7575 | 99.6150 |
|  | KS2 | 33.3415 | 33.5978 | 33.6319 | 33.5565 | 33.4891 | 33.5236 | 33.5236 | 33.5278 |
| AllBlack | KS1 | 99.6075 | 99.6175 | 99.6275 | 99.5850 | 99.6550 | 99.6250 | 99.6425 | 99.6075 |
|  | KS2 | 33.4289 | 33.5525 | 33.5671 | 33.4387 | 33.3660 | 33.3177 | 33.2725 | 33.3290 |
| AllWhite | KS1 | 99.6100 | 99.6300 | 99.5650 | 99.6150 | 99.6175 | 99.5375 | 99.6100 | 99.6100 |
|  | KS2 | 33.3634 | 33.6905 | 33.7103 | 33.4444 | 33.5743 | 33.4981 | 33.6602 | 33.3332 |

To observe the key sensitivity at the decryption level, we encrypt the plain image with a secret key and decrypt it with a slightly different secret key and then compare the correctness of the decrypted image with respect to the plain image by computing the perceptual metrics MAE, MSE and PSNR (already explained above). The strategy of a minor change in the secret key is the same as adopted above for the computation of KS1 and KS2. The results of our computation are summarized in Table 15. It is observed that the decryption with a slightly different key obtains a completely dissimilar image as compared to the plain image.

**Table 15:** Key sensitivity analysis results at the decryption level

|  |  | $X_0$ | $Y_0$ | $K$ | $N$ | $K_1$ | $K_2$ | $K_3$ | $K_4$ |
|---|---|---|---|---|---|---|---|---|---|
| Lena | MAE | 84.9971 | 85.3188 | 85.5766 | 84.3930 | 85.7021 | 85.2330 | 85.3190 | 85.0059 |
|  | MSE | 1.0847e+04 | 1.0897e+04 | 1.0957e+04 | 1.0766e+04 | 1.0987e+04 | 1.0879e+04 | 1.0886e+04 | 1.0854e+04 |
|  | PSNR | 7.7777 | 7.7577 | 7.7341 | 7.8103 | 7.7222 | 7.7650 | 7.7619 | 7.7748 |
| Baboon | MAE | 68.9074 | 69.4480 | 69.5254 | 69.3836 | 69.2587 | 69.6259 | 69.4769 | 69.1596 |
|  | MSE | 6.7418e+03 | 6.8477e+03 | 6.8487e+03 | 6.8201e+03 | 6.8425e+03 | 6.8810e+03 | 6.8572e+03 | 6.7934e+03 |
|  | PSNR | 9.8431 | 9.7754 | 9.7747 | 9.7929 | 9.7787 | 9.7543 | 9.7694 | 9.8099 |
| Peppers | MAE | 75.4327 | 75.0208 | 74.4845 | 74.9074 | 75.0093 | 75.2616 | 75.2777 | 75.0977 |
|  | MSE | 8.3501e+03 | 8.2700e+03 | 8.1647e+03 | 8.2750e+03 | 8.2714e+03 | 8.3492e+03 | 8.3180e+03 | 8.3262e+03 |
|  | PSNR | 8.9139 | 8.9557 | 9.0114 | 8.9531 | 8.9550 | 8.9143 | 8.9306 | 8.9264 |
| All Black | MAE | 127.5773 | 127.9102 | 127.7842 | 128.0183 | 127.3016 | 127.4850 | 127.8863 | 127.3536 |
|  | MSE | 2.1732e+04 | 2.1803e+04 | 2.1817e+04 | 2.1870e+04 | 2.1713e+04 | 2.1739e+04 | 2.1863e+04 | 2.1722e+04 |
|  | PSNR | 4.7599 | 4.7457 | 4.7428 | 4.7322 | 4.7635 | 4.7584 | 4.7336 | 4.7617 |
| All White | MAE | 127.6402 | 127.7985 | 127.6922 | 127.5577 | 128.0315 | 128.3397 | 126.9492 | 127.1920 |
|  | MSE | 2.1802e+04 | 2.1811e+04 | 2.1761e+04 | 2.1702e+04 | 2.1855e+04 | 2.1972e+04 | 2.1603e+04 | 2.1691e+04 |
|  | PSNR | 4.7458 | 4.7440 | 4.7539 | 4.7657 | 4.7353 | 4.7122 | 4.7858 | 4.7681 |

### 3.8 Key space analysis:

The secret key in the proposed image encryption algorithm is a set of one integer and seven floating-point numbers. The two floating-point numbers $(X_0, Y_0) \in (0, 2\pi)$ serve as the initial conditions for the chaotic conservative standard map, the remaining five floating-point numbers $(K, K_1, K_2, K_3, K_4) > 18.0$ serve as the parameter value for the conservative chaotic standard map at various stages of the algorithm and an integer $0 < N < 1000$ serves as the number of iterations to skip before using the map for the encryption purpose. The key sensitivity analysis reveals that the parameter and initial conditions of the conservative chaotic standard map differing by $10^{-14}$ can be treated as a distinct key. Since the initial conditions $(X_0, Y_0) \in (0, 2\pi)$ therefore there are $(2\pi \times 10^{-14})^2$ combinations of different keys for the Initial conditions. The parameter of standard chaotic map can have any value larger than 18.0 with a precision of $10^{-14}$ consequently, have infinite number of distinct combinations and there are $10^3$ different combinations for the value N. In this way, we may conclude that the proposed image encryption algorithm has infinite key space and consequently brute force attack is infeasible.

**3.8 Classical attack analysis:**

The most common and frequently used cryptanalytic attacks are known-plaintext attacks and chosen-plaintext attacks. In a known-plaintext attack the cryptanalyst knows the plaintext and its corresponding ciphertext and by establishing a meaningful relationship between the two along with the knowledge of the encryption algorithm tries to discover the secret key. In chosen plaintext attack, the cryptanalyst chooses multiple plaintexts of his/her choice (based on the intuition and knowledge of the structure of the encryption algorithm), generates the corresponding ciphertext for the same secret key (which is unknown) and then extracts some correlation, statistical information etc. to discover the secret key. The differential attack (see section 3.7) is also a kind of chosen-plaintext attack only. Sometimes adaptive chosen plaintext attacks are also implemented where one pair of plaintext (chosen in the first step) and corresponding ciphertext is analyzed and based on the results, the cryptanalyst chooses/creates a specific plaintext for the next step and further carried out the analysis and continues till the secret key is discovered.

Since the proposed image encryption algorithm exhibits extreme key, plaintext and ciphertext sensitivity (refer to Sections 3.7 and 3.8) therefore it is very difficult to extract any meaningful information through the pairs of plaintexts and cyphertexts. In the proposed image encryption algorithm, the dynamic one-time pixel (DOTP) is generated with the help of a part of the key and modified at the encryption of each pixel through the information from the plaintext and ciphertext generated so far. This DOTP is further used to encrypt the pixel and then the ciphertext generated so far will be used along with the plaintext information for the next DOTP modification and so on (i.e, feed -forward and feedback mechanisms). Moreover, the rules chosen for the encryption of each pixel are also dynamic and key-dependent. In the entire encryption process the secret key, plaintext and ciphertext are closely and sensitively interconnected such that an infinitesimal change in any of the components leads to a diverse effect in the resultant therefore implementation of any of the above-mentioned attacks appears completely infeasible.

## 4. Conclusion

The DNA encoding/decoding and operations (addition, subtraction, XOR, XNOR, complementing, etc.) if implemented jointly in a specific and strategic manner, under the control of chaotic systems, are capable of shuffling as well as altering the pixel values, therefore, may be effectively utilized for the image encryption. So far, many such algorithms have been developed and most of them are based on dissipative chaotic systems which possess the periodic windows and patterns in bifurcation diagrams, co-existing attractors in the neighborhoods of parameter space and are also characterized by the strange attractor which makes them prone to chaos-specific attacks and sometimes statistical attacks too. In this paper, we have proposed a novel conservative chaotic standard map-driven dynamic DNA coding (encoding, addition, subtraction and decoding) for the image encryption, which is the first (to the best of our knowledge) hybrid DNA and chaos-based image encryption based on conservative chaos. The algorithm also uses a novel method of generating pseudorandom sequences from the 2D conservative chaotic standard map which is validated for the pseudo randomness through NIST test suite before using it in the proposed algorithm.  A unique combination of feed-forward and feedback mechanisms has been incorporated along with a sequential inter-dependence (through the iterates of the chaotic map) while producing multiple pseudorandom sequences in the proposed image encryption algorithm to produce the desired plaintext, ciphertext and key sensitivities. The algorithm has been analyzed for its performance and security extensively through the most frequent, popular, contemporary and up-to-date quantitative metrics used in the field.  The results of our analysis are encouraging and prove the superiority, and robustness of the proposed algorithm against the most common cryptanalytic and statistical attacks.

.